\documentclass{aa}  
\usepackage{graphicx}
\usepackage{txfonts}
\usepackage[capitalise]{cleveref}
\usepackage{amsmath}	
\usepackage{amssymb,color}	

\newcommand{\vsini}{$v \sin i$}
\newcommand{\kms}{km\,s$^{-1}$}
\newcommand{\Rsun}{$R_{\odot}$}

\begin{document}

\title{The circumstellar environment of 55\,Cnc: }
\subtitle{The super-Earth 55\,Cnc\,e as a primary target for star-planet interactions\thanks{Based on observations obtained at the Telescope Bernard Lyot (USR5026) operated by the Observatoire Midi-Pyr\'en\'ees, Universit\'e de Toulouse (Paul Sabatier), Centre National de la Recherche Scientifique of France.}}
\titlerunning{55 Cancri}

\author{C. P. Folsom \inst{1}           
  \and
  D. \'{O}~Fionnag\'{a}in \inst{2}
  \and
  L. Fossati \inst{3}
  \and
  A. A. Vidotto \inst{2}
  \and
  C. Moutou \inst{1}
  \and
  P. Petit \inst{1}
  \and
  D. Dragomir \inst{4}
  \and
  J.-F. Donati \inst{1}
}

\institute{IRAP, Universit\'{e} de Toulouse, CNRS, UPS, CNES, 31400, Toulouse, France\\
  \email{cfolsom@irap.omp.eu}
  \and
  Trinity College Dublin, College Green, Dublin 2, Ireland
  \and
  Space Research Institute, Austrian Academy of Sciences, Schmiedlstrasse 6, A-8042 Graz, Austria
  \and
  Dept.~of Physics and Kavli Institute for Astrophysics and Space Research, MIT, Cambridge, MA 02139, USA and Hubble Fellow
}

\date{Received xxx; accepted yyy}

\abstract
   {55 Cancri hosts five known exoplanets, most notably the hot super-Earth 55\,Cnc\,e, which is one of the hottest known transiting super-Earths.} 
   {Due to the short orbital separation and host star brightness, 55\,Cnc\,e provides one of the best opportunities for studying \emph{star-planet interactions} (SPIs).  We aim to understand possible SPIs in this system, which requires a detailed understanding of the stellar magnetic field and wind impinging on the planet.}
   {Using spectropolarimetric observations, and Zeeman Doppler Imaging, we derive a map of the large-scale stellar magnetic field.  We then simulate the stellar wind starting from the magnetic field map, using a 3D MHD model. }
   {The map of the large-scale stellar magnetic field we derive has an average strength of 3.4\,G.  The field has a mostly dipolar geometry, with the dipole tilted by 90$^\circ$ with respect to the rotation axis, and dipolar strength of 5.8\,G at the magnetic pole.  The wind simulations based on this magnetic geometry lead us to conclude that 55\,Cnc\,e orbits inside the Alfv\'{e}n surface of the stellar wind, implying that effects from the planet on the wind can propagate back to the stellar surface and result in SPI. }
   {}

\keywords{stars: individual: 55 Cnc --
  stars: magnetic field --
  stars: late-type --
  stars: winds, outflows --
  stars: planetary systems --
  planet-star interactions
}

\maketitle
%

\section{Introduction}

The 55\,Cnc system is one of the most relevant  systems for understanding  planets with masses/radii in between those of Earth and Neptune.  This mass-radius regime, which is characterised by a large variety of bulk densities, is not found in the solar system, but yet constitutes the largest population of known exoplanets.  

55\,Cnc (G8V) hosts five known exoplanets, most notably the close-in super-Earth 55\,Cnc\,e (M$_p$\,=\,$8.37\,\pm\,0.38$\,M$_{\oplus}$, R$_p$\,=\,$1.92\,\pm\,0.08$\,R$_{\oplus}$, P$_{\rm orb}$\,=\,18\,h).  This offers one of the best opportunities for detailed characterisation of a super-Earth thanks to two key properties: at V$\approx$6, the host star is the third brightest star known to host a transiting planet, and because of the high planetary temperature, there is a high planet-star flux contrast, making the planetary emission well detectable at optical and infrared (IR) wavelengths.

Observations suggest that 55\,Cnc and planet e may be an up-scaled analogue of the Jupiter-Io system \citep{Demory2016}. This would imply the presence of a significant exosphere surrounding the planet and plasma flowing from the planet towards the star, following the stellar magnetic field lines. Spitzer observations conducted at 4.5-$\mu$m indicate the planet's dayside thermal emission varied by about 300\% between 2012 and 2013, with temperatures ranging between 1300 and 3000\,K \citep{Demory2016b}. This can be explained assuming that the planetary lithosphere is partially molten, particularly on the dayside. 

MOST satellite observations \citep{Winn2011,Dragomir2014} show variability in the visible that could be linked to the IR variability. The MOST phase-curve in 2012 shows a smaller amplitude than in 2011, and the former is comparable to the IR phase curve.  This has lead to suggestions that there is material opaque both in the visible and in the IR, such as grains of dust coupled to hot plasma, similar to Io \citep{Kruger2003}. The grey behaviour of dust could provide the same opacity both in the visible and IR. 

More recently, \citet{Ridden-Harper2016} detected absorption at the position of the Ca{\sc ii}\,H\&K resonance lines, which may be connected with the exosphere of planet e. This 4.9$\sigma$ detection was achieved only for one of the 4 datasets, suggesting temporal variability in the optical depth of the material surrounding the planet.

Reconciling these properties would require the presence of azimuthally inhomogeneous circumstellar material and/or of a large exosphere made of ions and charged dust particles similar to Io \citep{Kruger2003} that could contribute to a variable grey opacity along the line of sight through diffusion of stellar light. 
The large scale structure of the circumstellar material appears to remain stable for days or weeks \citep{Demory2016b}, which would likely require a large scale stellar magnetic field for support.

A large stellar magnetic field would also be able to connect star and planet, creating chromospheric hotspots. Again, we find an analogy with the Io-Jupiter system, in which observations indicate the existence of auroral hotspots in Jupiter at the footpoints of magnetic field lines connecting Io to Jupiter \citep{2000RvGeo..38..295B}. Therefore, detecting the presence of a significant magnetic field on 55 Cnc would support the idea that this system is a scaled-up version of the Jupiter-Io system. Examining the stellar magnetic and wind properties would also clarify the likelihood of magnetic star-planet interactions.

We present the detection and characterisation of the stellar surface magnetic field of 55 Cnc and model its stellar wind. This paper follows an earlier unsuccessful search for a magnetic field for 55 Cnc by \citet{marcy1984} using Zeeman broadening. From our simulations, we propose that 55 Cnc e should be a primary target for detecting star-planet interactions, and instrumentation available in the near future may indeed lead to such a detection.

\section{Zeeman Doppler Imaging}
\label{sec:zdi}
To detect and characterise the magnetic field of 55 Cnc, we observed the star using the Narval spectropolarimeter at the Pic du Midi Observatory in France.  Narval includes a polarimeter module connected by fibre to a high resolution (R$\sim$65000) \'echelle spectrograph (3700 to 10500 \AA ). Observations were made in Stokes $V$ mode, providing both total intensity (Stokes $I$) and circularly polarised (Stokes $V$) spectra.  Data were reduced using the {\sc LibreESPRIT} pipeline \citep{Donati1997-major} in the version built for Narval.  

We obtained 20 observations of 55 Cnc on 10 nights between March 1st and April 21st 2017.  Two consecutive observations were obtained on each night to allow the possibility of co-adding observations, however since magnetic detections were obtained reliably in single observations co-adding was not needed.  A total exposure time of 3600s (as a sequence of four 900s sub-exposures) was used.  

\begin{figure}
    \centering
    \includegraphics[width=1.0\linewidth]{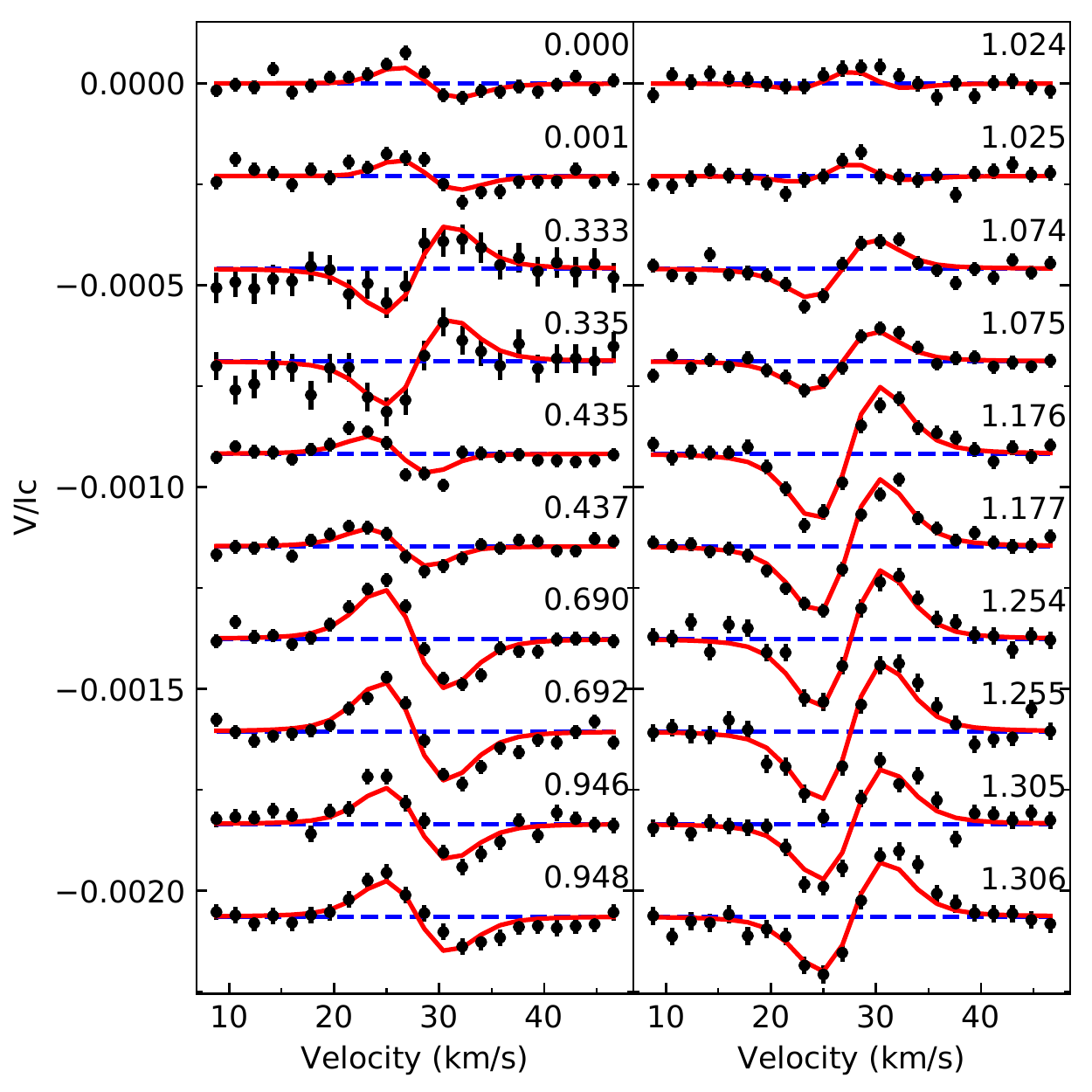}
    \vspace{-0.5cm}
    \caption{LSD Stokes $V$ profiles of 55 Cnc (black points), labelled by rotation cycle for a 39-day period, and shifted vertically for clarity.  Dashed lines indicate zero and solid lines are the best fit ZDI model profiles.}
    \label{fig:prof-V-fit}
\end{figure}

To detect Zeeman splitting in Stokes $V$, we used Least Squares Deconvolution \citep[LSD,][]{Donati1997-major, Kochukhov2010-LSD}.  This is a cross-correlation technique that produces pseudo-average line profiles with greatly reduced noise.  We used the line mask for a 5000K star from \citet{Folsom2018-Toupies2}, a normalising wavelength of 650nm, a normalising Land\'e factor of 1.195, and the code of \citet{Donati1997-major}. We obtained magnetic detections in the $V$ LSD profiles (\cref{fig:prof-V-fit}) on all nights except for April 10, and those non-detections appear to be due to the rotation phase of the star.  

We characterised the magnetic field strength and geometry of 55 Cnc using Zeeman Doppler Imaging (ZDI, e.g.\ \citealt{Donati2006-ZDI-SpH}).  ZDI uses a regularised fitting procedure to model the rotationally modulated Stokes $V$ LSD profiles and infer the simplest magnetic geometry needed to reproduce them. We used the ZDI code of \citet{Folsom2018-Toupies2}, following the analysis procedure of \citet{Folsom2018-HD219134}.  The \vsini\ of 55 Cnc is poorly constrained since it is below the resolution of Narval.  We inferred an equatorial velocity of 1.24 \kms, using the interferometric radius of $0.943 \pm 0.010$ \Rsun\ from \citet{vonBraun2011-55Cnc-phys-param} (confirmed by \citealt{Ligi2016-phys-param}), and the rotation period of $38.8 \pm 0.05$ days from \citet{Bourrier2018} (within $3\sigma$ of $40.7 \pm 0.7$ days from \citealt{Hempelmann2016-CaHK-periods}).  To constrain the intrinsic line profile shape we fit the line in Stokes $I$ with a Voigt profile, broadened by this upper limit on \vsini, and found a Gaussian width of 2.32 \kms\ and a Lorentzian width of 0.81 (2.84 \kms) (c.f. \citealt{Folsom2018-HD219134}). 

Using the above equatorial velocity and intrinsic line widths, we constrain the inclination of the stellar rotation axis ($i$) using ZDI to fit the Stokes $V$ profile variability.  This was done by fitting ZDI models with a wide range of inclinations and looking for values that provided the best fit.  We find $i > 80^\circ$ with a $1\sigma$ confidence (and $i > 65^\circ$ at a $3\sigma$ confidence), with the most likely values near $90^\circ$.   \citet{Bourrier2018} found an orbital inclination for 55 Cnc e of $83.59^{+0.47}_{-0.44}$$^\circ$, thus our rotation inclination is consistent with the stellar rotation axis and planet orbital axis being aligned.

We re-derived the rotation period using ZDI fitting of the LSD profiles, as was done by \citet{Folsom2018-HD219134}, and found $39.0 \pm 0.3$ days, in good agreement with \citet{Bourrier2018}.  We searched for surface differential rotation using ZDI, as in \citet{Folsom2018-HD219134}, but found no strong evidence for it. 
We found that $d\Omega$ (the difference in angular frequency between the pole and equator) is $< 0.065$ rad\,day$^{-1}$ at a $3\sigma$ confidence level, and that this parameter covaries strongly with rotation period.  The uncertainty is based on variations in $\chi^2$ for models computed with different period and $d\Omega$ but the same entropy, as in \citet{Folsom2018-HD219134}. This result implies that differential rotation, if at all present, is not anomalously strong in this star, compared to other stars of this spectral type. The specific value of $d\Omega$ within this uncertainty range has a minimal impact on the magnetic map, changing field strengths and how poloidal or axisymmetric the field is by less than 2\%. Our data span only 1.3 rotation cycles, further complicated by the low \vsini, which likely explains the lack of evidence for differential rotation in the data.

\begin{figure}
    \centering
    \includegraphics[width=1.0\linewidth]{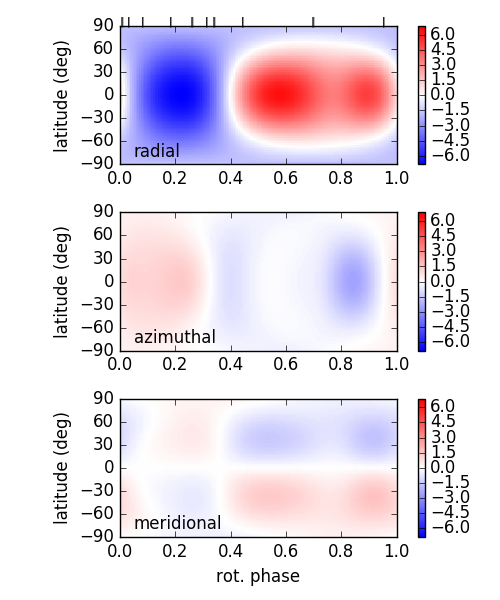}
    \vspace{-0.5cm}
    \caption{Magnetic map of 55 Cnc from ZDI. The radial, azimuthal, and meridional components of the magnetic field are presented, in units of G.  Ticks above the top panel indicate phases where observations were obtained.}
    \label{fig:zdi-map}
\end{figure}

The final magnetic map is presented in Fig.\ \ref{fig:zdi-map}, and corresponding line profile fits are in Fig.\ \ref{fig:prof-V-fit}, for $i=90^{\circ}$, $P = 39.0$ days, and $d\Omega = 0$ rad\,day$^{-1}$.  We find an average unsigned large-scale field of 3.4 G.  The large-scale field we reconstruct is dominantly poloidal (99\% of the magnetic energy, as calculated from $B^2$), and most of that poloidal field is in the dipole (79\% energy) and quadrupole (19\% energy) components.  The strength of the dipolar component, at the magnetic pole, is 5.8 G. The magnetic field we find is dominantly non-axisymmetric (94\% energy, $m \neq 0$), although that may be biased by the stellar inclination being near $90^\circ$.

To investigate the impact of an unexpectedly large error in the inclination, we computed a magnetic map with $i = 60$, and found the map was largely unchanged.  The axisymmetry of the poloidal component is virtually unchanged, while the dipolar axisymmetry increases by 5\%.  We derive the same poloidal fraction, although 4\% of the poloidal energy shifts from the dipole to the quadrupole and octopole components.

As the inclination approaches $90^\circ$, ZDI (like regular Doppler Imaging) suffers from north-south `mirroring' effects.  More precisely, in terms of spherical harmonics, components where $l+m$ is an odd number, i.e.\ where $B(\theta,\phi) = -B(\pi-\theta,\phi)$, are not constrained when $i=90^\circ$ and so are forced to zero by the regularisation. Most notably that includes the axisymmetric dipole component ($l=1, m=0$), so there is likely some magnetic field that our observations are not sensitive to.  The phase sampling of the observations between 0.5 and 1.0 is relatively coarse, which likely lowers the resolution of the magnetic map in that hemisphere. However, for the stellar wind near planet e and beyond, only the most large scale features (lowest degree spherical harmonic) dominate \citep[e.g.][]{Jardine2017}.

\section{The stellar wind around 55 Cnc}
\label{sec:planets}
We used the numerical modelling tool BATS-R-US \citep{Powell1999} to simulate the stellar wind of 55 Cnc. This code has been used to simulate astrophysical plasma environments \citep{Toth2005,ManchesterIV2008a,Vidotto2015,Alvarado-Gomez2018}, and here we used the version from \citet[][more details can be found in that paper]{Vidotto2015}. BATS-R-US solves the set of closed, ideal, MHD equations for mass, momentum and energy conservation, and the magnetic induction equation.  The code solves for 8 parameters: mass density ($\rho$), velocity ($\textbf{u} = \lbrace u_x, u_y, u_z \rbrace$), magnetic field ($\textbf{B} = \lbrace B_x, B_y, B_z \rbrace $), and gas pressure (P). We assume that the plasma behaves as an ideal gas, hence $P = n k_B T $, where $n = \rho / (\mu m_p)$ is the total number density of the wind, with $\mu m_p$ denoting the average particle mass. We take $\mu = 0.5$, which represents a fully ionised hydrogen wind. Pressure is related to density in the wind by the polytropic relation: $P \propto \rho^{\gamma}$, where $\gamma$ is the polytropic index. Through the polytropic index, heating is implicitly added to the wind. We adopt $\gamma = 1.05$, similar to values used in the literature \citep{Vidotto2015,Reville2015,OFionnagain2018}. Including the polytropic index, this model contains three free parameters, with the remaining two being the base density ($\rho_0$) and base temperature ($T_0$) of the wind. For this simulation we use $\rho_0 = 10^{-16}$ g cm$^{-3}$ and $T_0 = 1$ MK. We simulate the stellar wind in a grid that extends to 30 R$_{\star}$ in each direction.  Our simulation has a range of resolutions from 0.019 -- 0.625 R$_{\star}$. We find that 55 Cnc is losing mass at a rate of $2.2 \times 10^{-14}$ M$_{\odot}$ yr$^{-1}$, similar to the solar mass-loss rate of $2 \times 10^{-14}$ M$_{\odot}$ yr$^{-1}$, and angular momentum at a rate of $8.2 \times 10^{29}$ erg. 

\begin{figure}
    \centering
    \includegraphics[width=.47\textwidth]{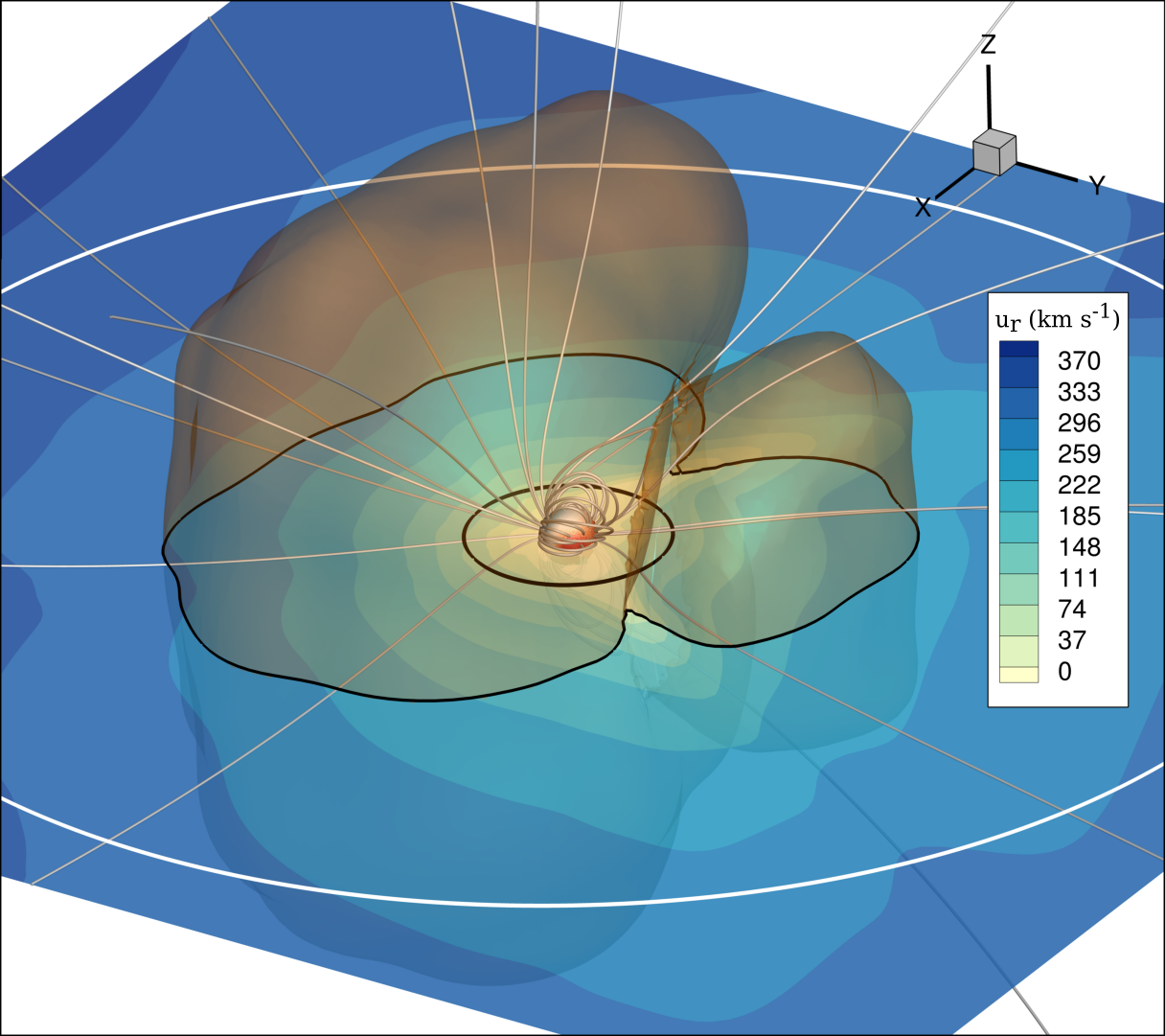}
    \caption{3D MHD simulation of 55 Cnc. Wind velocity is shown as a slice at the z = 0 plane and extends to 30 R$_{\star}$. The Alfv\'{e}n surface is shown in orange, with its intersection with the orbital plane of the planets highlighted with a thick black line. Open magnetic field lines are shown in grey and closed magnetic field lines are shown in red. The orbits of 55 Cnc e and b are displayed in black and white respectively.\label{fig:55cnc-3d}}
\end{figure}

\Cref{fig:55cnc-3d} shows the output of our simulation, where the black and white circles represent the orbits of planets e and b, at 3.5 and 26.2 R$_{\star}$, respectively.  Extracting values from the wind at these orbits allows us to investigate the environment surrounding these planets. \Cref{table1} shows the values of the stellar wind properties averaged over one planetary orbit for each of the known planets. As some of the planets are outside the domain of our simulation, we perform an extrapolation of the wind values at planets c, f, and d.  While planet e is impacted by a much slower wind, the wind still imparts a higher ram pressure upon planet e than upon planet b, due to the much higher wind density closer to the star. 

\Cref{fig:55cnc-3d} shows that planet e orbits entirely within the Alfv\'{e}n surface of the wind of 55 Cnc. This means that the wind is magnetically dominated around this planet. 
Being in a sub-Alfv\'{e}nic regime means that planet e has a direct connection to the star -- in the sub-Alfvenic regime, the planet can generate Alfven wings, which can couple to the star and carry electromagnetic energy toward the star \citep[e.g.][]{2013A&A...552A.119S, 2019ApJ...872..113F}. This scenario has ramifications for the star-planet interactions (SPI), as we will discuss in the next section. An important factor to keep in mind is that the size of the Alfv\'{e}n surface depends on the stellar magnetic field and wind base density. Although the stellar magnetic field is tied to the observations, the density is a free parameter in our model. Thus, we performed two additional sets of simulations, with 3 times larger and 3 times smaller base density.  
These choices affect, for example, the mass-loss rate of the stellar wind, but in all three scenarios we explored, the orbit of planet e remains entirely within the Alfv\'{e}n surface.

\begin{table*}
\begin{center}
\caption{Stellar wind local properties around the known 55 Cnc planets averaged over one planetary orbit.} \label{table1} 
\begin{tabular}{lcccccccccc}
\hline
Orbital Wind Properties &  `e' &  `b'& `c' & `f' & `d' & Reference  \\
\hline\hline
orbital period (days) & $0.74$ &  $14.7$ & $44.4$ & $262$ & $4825$ & \citet{Baluev2015}\\ 
semi-major axis (au) &  $0.0154$ & $0.1134$ & $0.2373$
 & $0.7708$ & $5.957$ & \citet{Bourrier2018} \\ 
semi-major axis $(R_\star)$ & 3.52 & 25.85 & 54.10 & 175.74 & 1358.17 & \citet{Bourrier2018} \\ 
stellar wind density (g~cm$^{-3}$) & $4.5\times10^{-19}$ & $7.7\times10^{-22}$ & $1.4\times10^{-22}$ & $1.2\times10^{-23}$ & $1.7\times10^{-25}$ & this work \\
 velocity (km~s$^{-1}$) & 87.5 & 343.0 & 384.5 & 384.5 & 384.5 &  this work \\
velocity incl.~orbital motion (km~s$^{-1}$) &  244.1 & 349.4 & 388.9 & 385.9 & 384.6 &  this work \\
ram pressure ($10^{-7}$\,dyn cm$^{-2}$) & 313.1 & 8.69 & 2.06 & 0.27 & 0.008 &  this work\\
 temperature ($10^{6}$ K)  & 1.08 & 0.837 & 0.761 & 0.666 & 0.528 &  this work\\
 magnetic field  (nT) & 5021 & 39.02 & 7.04 & 0.56 & 0.007 &  this work\\
\hline
\end{tabular}
\end{center}
\end{table*}

\section{SPI induced by planet 55 Cnc e}
\begin{figure}
    \centering
    \includegraphics[width=\linewidth]{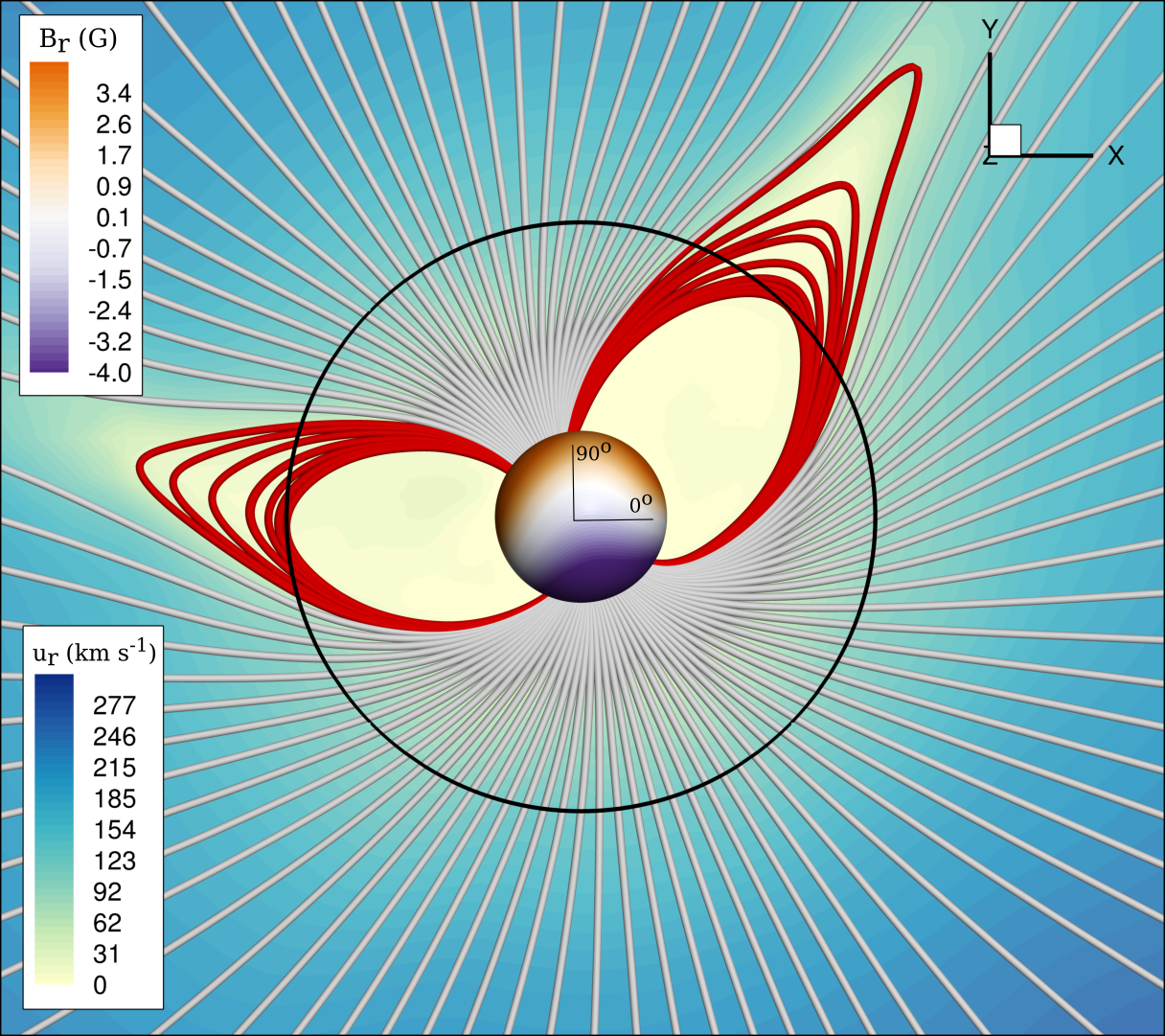}
    \caption{Stellar magnetic field lines that intercept planet e as it orbits around the star (black circle). Stellar rotation axis is perpendicular to this orbital plane. The red field lines  correspond to closed field regions. The top colourbar shows  magnetic field strength (purple-orange) and bottom one shows the stellar wind velocity (yellow-blue). }
    \label{fig:SPI_Blines}
\end{figure}

It has been suggested that close-in planets could enhance stellar activity via SPI, which would be able to generate induced-chromospheric hotspots in the host star \citep{Cuntz2000, Cuntz2002}. Potential signatures of SPI in the form of anomalous chromospheric emission modulated by a planet has been reported for a few systems, such as in the case of HD179949 \citep{Shkolnik2008, Fares2012} and HD189733 \citep{2018AJ....156..262C}. However, not all close-in planets can generate induced emission in their host stars \citep{Shkolnik2005,Shkolnik2008,Miller2012,miller2015}. One possible suggestion is that if the planet orbits beyond the Alfv\'en surface, information from a potential star-planet magnetic reconnection event would be prevented from propagating towards the star. Given the sub-Alfv\'{e}nic orbit of 55 Cnc e, we investigate here the potential sites of generation of chromospheric hotspots, caused by SPI. \Cref{fig:SPI_Blines} shows a set of 100 magnetic field lines that intercept planet e as it orbits around the star.  If chromospheric hotspots can be formed due to ``magnetic SPI'' (i.e., through magnetic reconnection or Alfv\'{e}n wings; \citealt{Neubauer1980}, \citealt{2004ApJ...602L..53I}), the footpoints of the connecting lines would tell us where and when such hotspots would appear.

\begin{figure}
  \centering
  \includegraphics[width=\linewidth]{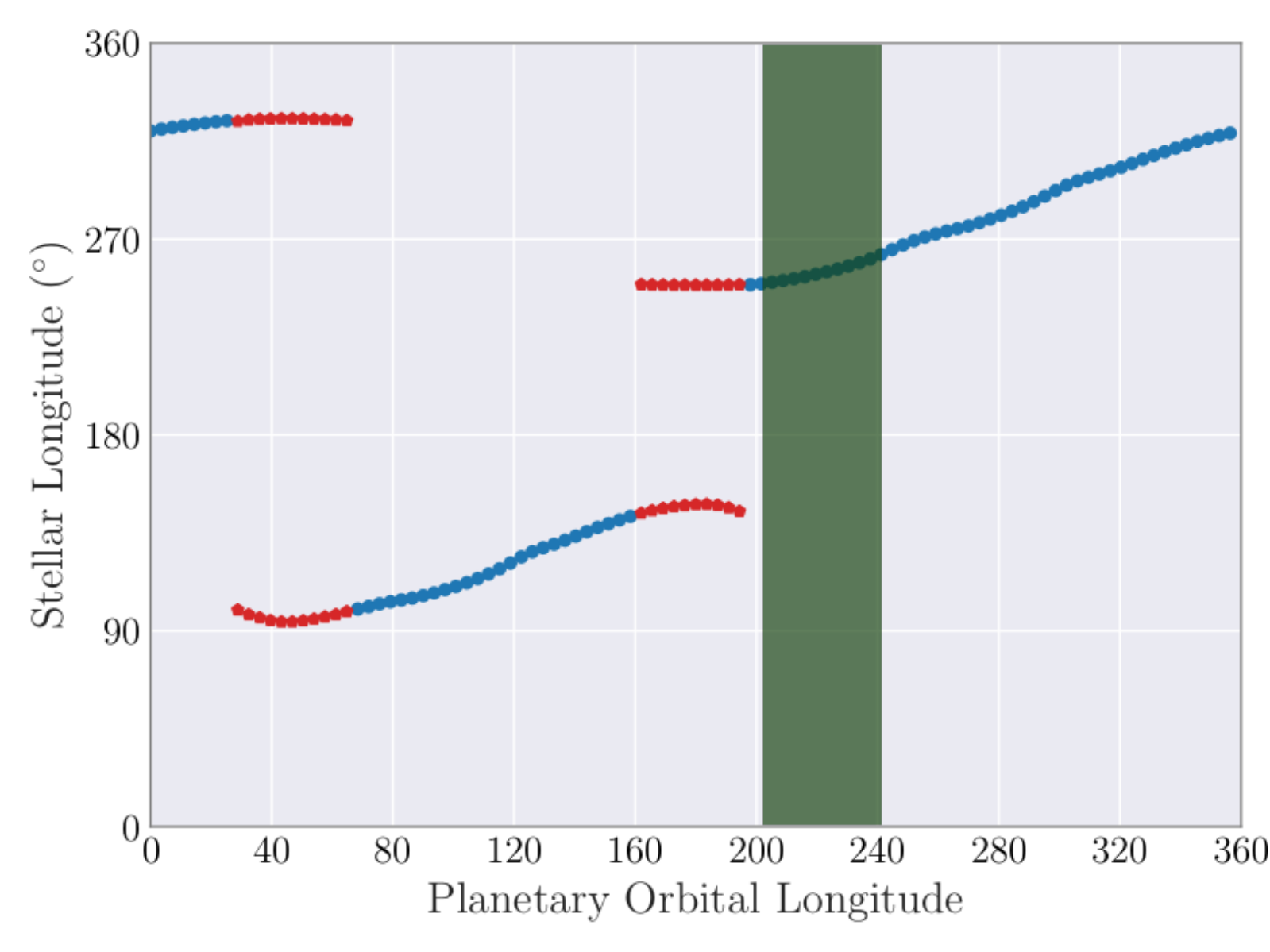}
  \includegraphics[width=\linewidth]{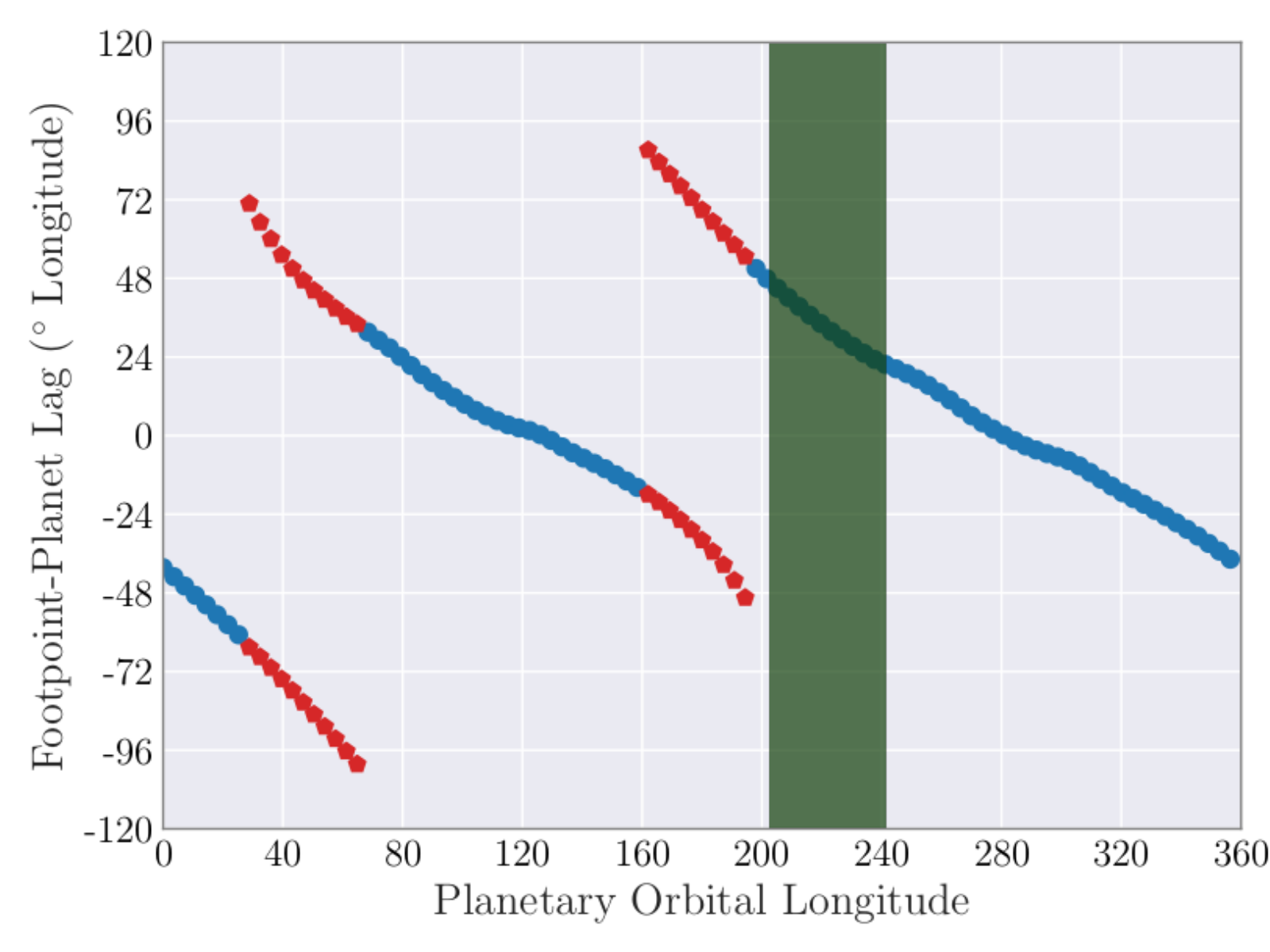}
  \vspace{-0.5cm}
  \caption{\emph{Top:} The longitude of the footpoint on the stellar surface vs.\ the longitude of the planetary orbit, as defined in \cref{fig:SPI_Blines}. Footpoints of open (closed) magnetic field lines are shown in blue (red).  \emph{Bottom:} Phase lag (difference in longitudes between footpoint and subplanetary point) versus the orbital longitude of the planet. The footpoints move from lagging behind the planet orbit to ahead of it. 
    For the closed loops, it is expected that the SPI will occur at both footpoints of the loop, unless that loop extends beyond the Alfv\'{e}n surface. 
    The green region indicates the location during planetary transit.}
  \label{fig:footpoints_image}
\end{figure}

\Cref{fig:SPI_Blines} shows that the magnetic field lines linking the planet to the star are a combination of open lines and closed loops. These closed magnetic loops will cause the SPI-related chromospheric spots on the star to move differently from the planetary orbit, with large jumps occurring where the planet moves from one branch of the closed loop to another. This phase lag and jumping effect is evident from \cref{fig:footpoints_image} (see also predictions by \citealt{Mcivor2006}). Recently, \citet{Strugarek2019} showed a similar effect in the case of Kepler-78, where the magnetic topology of the host star can greatly affect the transient nature of SPI. 
Although their simulations did not explain the amplitude of enhanced activity observed by \citet{moutou2016}, in stars with stronger magnetic fields \citep[e.g.\ HD 179949, ][]{Fares2012} the effect may be more detectable.

As the observed magnetic field of 55 Cnc is largely described by a dipole tilted by 90$^\circ$ to the rotation axis, all of the magnetic field lines that intersect with the planetary orbit lie within the equatorial plane. This is specific for the case of 55 Cnc and specific to the epoch of ZDI observation. For other planetary systems and/or different observing epochs, the magnetic field footpoints may lie above/below the equator, and this is  dependent on the large-scale stellar magnetic field geometry. 
The magnetic field is essential for identifying the theoretical period of modulation for magnetic SPI.  An aligned dipole will generate a modulation at the synodic period (0.73 days for 55 Cnc e), while a dipole tilted at 90$^\circ$ will generate a modulation of half the synodic period. The visibility of magnetic footpoints will also be modulated by the stellar rotation period. For 55 Cnc, because the field is not purely dipolar, the distribution of footpoints on the stellar surface is not symmetric about the rotation axis -- the branch of footpoints of positive polarity (\cref{fig:SPI_Blines}) occupies a smaller range of stellar longitudes than that of negative polarities. 
This can also be seen in the top panel of \cref{fig:footpoints_image}, where we see that the branch of footpoints above 180$^\circ$ of stellar longitude lasts longer than the footpoints below 180$^\circ$. An additional phase lag between the planet and induced stellar activity may be present due to the time it takes the Alfv\'{e}n waves to travel to the star \citep{Strugarek2019}. For 55 Cnc, though, this travel time  lag is less than 10\% of the orbital period.  

Due to the inclination of the rotation axis of 55 Cnc being near $90^\circ$, we cannot strongly constrain the axisymmetric dipolar component of the magnetic field.  Unfortunately, this limitation is intrinsic to any Stokes $V$ observations, and additional observations will not simply resolve it.  Depending on how strong an unseen $m=0$ dipole is, this could change the position of hotspots on the surface of the star, mostly in latitude, as the connectivity between the stellar magnetic field and planet could change. In terms of wind modelling, adding a dipolar component with $m=0$ to the magnetic map of 55 Cnc derived here would imply an Alfv\'{e}n surface that is larger than the one we computed. In this case, an unseen axisymmetric dipolar component would only reinforce our results that 55 Cnc e orbits inside the Alfv\'{e}n surface, strengthening possible SPI. 

\section{Summary and Discussion}
To summarise, we have used spectropolarimetric observations of 55 Cnc to detect the stellar magnetic field, and mapped the large-scale magnetic field using ZDI, finding largely a tiled dipolar field. This magnetic map was used as input for a 3D MHD simulation of the stellar wind, and the wind's properties at the positions of the known planets were estimated. We found that 55 Cnc e orbits entirely within the Alfv\'{e}n surface of the stellar wind, which implies magnetic field lines connect planet e to the star, allowing for magnetic star-planet interactions. Using these simulations, we estimated the possible position of chromospheric hotspots due to this interaction, and found they would be offset in phase from the planet's orbital position, and that apparent activity due to SPI may be modulated with a period close to twice the orbital or synodic period. Recently, \citet{sulis2019} detected a stellar flux modulation in phase with the orbital period of planet e. Furthermore, the amplitude of this modulation appears to be directly related to stellar activity, indicating that this could be caused by  magnetic interaction occurring as a consequence of the planet lying entirely within the Alfv\'{e}n surface of the star. Assuming the interpretation of the observation by \citet{sulis2019} in terms of star-planet interaction is correct, the possible apparent discrepancy in the timing of the modulation between the observation and the model prediction may be caused by the fact that the photometric and spectropolarimetric observations have been obtained more than two years apart: within this time, the geometry and strength of the stellar magnetic field may have changed.

Ideally, contemporaneous observations of different activity proxies would be able to better characterise any signature of star-planet interaction. Additionally, future spectropolarimetric observations of the system would also allow us to investigate whether the large-scale field is varying with time, and how. In particular, we would be able to further investigate the axisymmetric component, as here we found that the magnetic field is dominantly non-axisymmetric. New spectropolarimetric observations over a longer time period would also allow us to further constrain the level of differential rotation on 55 Cnc. In our analysis, it appears that differential rotation is not exceptionally strong in this star. Thanks to its high photometric precision, the CHEOPS mission will be able to follow the amplitude of the photometric modulation, which would be particularly powerful when combined with contemporaneous spectropolarimetric monitoring  \citep[e.g.][]{Fares2017}.

\begin{acknowledgements}
  AAV, D\'{O}F, CPF acknowledge joint funding received from the Irish Research Council and Campus France through the Ulysses funding scheme. The authors acknowledge the DJEI/DES/SFI/HEA Irish Centre for High-End Computing (ICHEC) for the provision of computational facilities and support. This work used the BATS-R-US tools developed at the University of Michigan Center for Space Environment Modeling and made available through the NASA Community Coordinated Modeling Center. D\'{O}F acknowledges funding received from the Trinity College Postgraduate Award. DD acknowledges NASA Hubble Fellowship (HST-HF2-51372.001-A). AAV acknowledges funding received from the Irish Research Council Laureate Awards 2017/2018 and from the European Research Council (ERC) under the European Union's Horizon 2020 research and innovation programme (grant agreement No 817540, ASTROFLOW). JFD acknowledges funding from the European Research Council (ERC) under the H2020 research \& innovation programme (grant agreement \#740651 NewWorlds). 
\end{acknowledgements}

%
%
\bibliographystyle{aa}
\bibliography{library}

\end{document}